\documentclass[10pt]{article}

 \usepackage{amsmath,amsthm,amssymb}
 \usepackage{makeidx,epsfig,lscape}
 \usepackage{color,colortbl}
 \usepackage{fancyhdr}
 \usepackage{xcolor,pict2e}

 \renewcommand{\headrulewidth}{0pt}
 \renewcommand{\footrulewidth}{0.5pt}

 \definecolor{myaqua}{rgb}{0.0,0.5,0.55}
 \definecolor{lightaqua}{rgb}{0.75,0.95,0.95}

\def\lin#1#2{\textcolor[rgb]{0.6,0.6,0.6}{\vspace*{#1mm} \hrule
   height 3 pt \vspace*{#2mm}}}

\def\bt{\begin{tabular}}
\def\et{\end{tabular}}
\def\and{\mbox{ and }}

\def\1{{\bf 1}}

 \def\sectionn#1{\refstepcounter{section}{\color{myaqua}

 \vskip 6mm

 \noindent\Large\bf\thesection. #1}

 \vskip 3mm}
 \def\subsectionn#1{\refstepcounter{subsection}{\color{myaqua}

 \vskip 5mm

 \noindent\large\bf\thesubsection. #1}

 \vskip 2mm}

 \def\boxx#1#2#3#4#5{
 {\linethickness{#4pt}\put(#1,#5){\color{myaqua}{\line(1,0){#3}}}}
 \multiput(#1,#2)(0,#4){2}{\line(1,0){#3}}
 \multiput(#1,#2)(#3,0){2}{\line(0,1){#4}}
  }
\usepackage{graphicx}
\usepackage{subfigure}
\usepackage{bm}
\newcommand{\mcl}{\mathcal}
\newcommand{\Slash}[1]{{\ooalign{\hfil/\hfil\crcr$#1$}}}
\newcommand{\nn}{\nonumber\\}
\newcommand{\vev}[1]{\langle #1 \rangle}  
\newcommand{\gr}[2]{\text{#1}(#2)}
\graphicspath{{./figs/}}

\begin{document}

 \fancyhead[L]{\hspace*{-13mm}
 \bt{l}{\bf Journal of Modern Physics, 2016}\\
 Published Online April 2016 in SciRes.
{\color{blue}{\underline{\smash{http://www.scirp.org/journal/jmp}}}} \\
 \et}
 \fancyhead[R]{\includegraphics{pic1.ps}}

 $\mbox{ }$

 \vskip 12mm

{ 
{\noindent{\huge\bf\color{myaqua}
Non-perturbative Analysis of Various Mass \\[2mm] Generation 
by Gluonic Dressing Effect with \\[2mm] the Schwinger-Dyson Formalism in QCD
}}
\\[6mm]
{\large\bf Shotaro Imai$^1$, Hideo Suganum$^2$}}
\\[2mm]
{
\noindent
$^1$ 
  Institute for the Advancement of Higher Education,
  Hokkaido University, Sapporo 060-0810, Japan 
\\
Email:\smash{imai@nucl.sci.hokudai.ac.jp}
\\[1mm]
$^2$Department of Physics, Graduate School of Science,
  Kyoto University, Sakyo, Kyoto 606-8502, Japan\\
Email:\smash{suganuma@scphys.kyoto-u.ac.jp}
 \\[4mm]

 \includegraphics{pic2.ps}

\lin{5}{7}

 { 
 {\noindent{\large\bf\color{myaqua} Abstract}{\bf \\[3mm]
 \textup{
As a topic of ``quantum color dynamics'', we study 
various mass generation of colored particles and gluonic dressing effect 
in a non-perturbative manner, 
using the Schwinger-Dyson (SD) formalism in (scalar) QCD.
First, we review dynamical quark-mass generation in QCD in the SD approach 
as a typical fermion-mass generation via spontaneous chiral-symmetry breaking. 
Second, using the SD formalism for scalar QCD, 
we investigate the scalar diquark, a bound-state-like object of two quarks, 
and its mass generation, which is clearly non-chiral-origin. 
Here, the scalar diquark is treated as an extended colored scalar field, 
like a meson in effective hadron models, 
and its effective size $R$ is introduced as a form factor. 
As a diagrammatical difference, 
the SD equation for the scalar diquark has an 
additional 4-point interaction term, in comparison with the single quark case. 
The diquark size $R$ is taken to be smaller than a hadron, 
$R\sim 1$ fm, and larger than a constituent quark, $R\sim 0.3$ fm. 
We find that the compact diquark with $R\simeq 0.3$ fm has a large effective 
mass of about 900 MeV, and therefore such a compact 
diquark is not acceptable in effective models for hadrons. 
We also consider the artificial removal of 3- and 4-point interaction, 
respectively, to see the role of each term, and find that the 4-point 
interaction plays the dominant role of the diquark self-energy.
From the above two different cases, quarks and diquarks, we guess that 
the mass generation of colored particles is 
a general result of non-perturbative gluonic dressing effect.
 }}}
 \\[4mm]
 {\noindent{\large\bf\color{myaqua} Keywords}{\bf \\[3mm]
Dynamical Mass Generation; Diquarks; Schwinger-Dyson Formalism; QCD
}

\lin{3}{1}

\newpage

\sectionn{Introduction}

{ \fontfamily{times}\selectfont
  \noindent

Quantum chromodynamics (QCD) is the fundamental gauge theory of the strong 
interaction, and it is a long important problem to describe 
hadron structure and properties based on QCD. 
Quarks and gluons, the basic ingredients of QCD, strongly interact with 
each other in an infrared region, and they are confined in hadrons. 
Then, due to their non-perturbative properties, it is fairly difficult 
to describe hadrons directly from QCD.
Also, the non-perturbative dynamics in QCD directly 
relates to the other important physical subject of ``mass generation.''

The origin of mass is one of the most fundamental issues in physics. 
One famous category of mass generation is the Yukawa interaction 
with the Higgs field. 
However, even besides the dark sector, 
the Higgs-origin mass is only about 1\% 
of the total mass in our universe, where dominant massive particles are 
nuclei (u,d quarks) and electrons. 
Actually, the Higgs interaction only gives 
the electron mass (about 0.5MeV) and 
a small current quark mass (a few MeV) for u,d quarks~\cite{PDG}. 
In contrast, about 99\% of mass of matter in our universe 
are created by the strong interaction, 
apart from the dark sector. 
In fact, a large constituent quark mass of 
$M_\psi =(300-400){\rm MeV}$ arises from non-perturbative dynamics in QCD. 
Thus, QCD gives another category of mass generation.

Such a dynamical fermion-mass generation in the strong interaction 
was first pointed out by Y.~Nambu~et~al. \cite{Nambu1961} in 1961 
in the context of spontaneous chiral-symmetry breaking. 
The QCD-based quantitative analysis of dynamical fermion mass generation 
was performed by Higashijima and Miransky 
in 1980's \cite{Miransky199402, Higashijima1984}
using the Schwinger-Dyson formalism. 
Thus, light u,d-quarks are considered to acquire 
a large constituent quark mass of about $300-400{\rm MeV}$, 
in accordance with spontaneous chiral-symmetry breaking.

Even without chiral symmetry breaking, however, it is likely that 
QCD has several dynamical mass generation mechanism. 
For example, while the charm quark has no chiral symmetry, 
some difference seems to appear between current and 
constituent masses for charm quarks: the current mass is $m_c\simeq 1.2$ GeV 
at renormalization point $\mu=2$ GeV~\cite{PDG}, and the constituent charm 
quark mass is $M_c\simeq 1.6$ GeV in the quark model.
The gluon is more drastic case. 
While the gluon mass is zero in perturbation QCD, 
the non-perturbative effect of the self-interaction of gluons 
seems to generate a large effective mass of 
0.6 GeV~\cite{Iritani2009,Mandula1987,Amemiya1999}, 
and the lowest glueball mass is about 1.6GeV~\cite{Morningstar1999,Ishii2002}. 
Furthermore, the dynamical mass generation for scalar-quark have been studied 
in the lattice scalar-QCD calculation~\cite{Iida2007}.
Thus, we deduce that ``quantum color dynamics'' generally 
accompanies a large mass generation, due to the strong interaction. 

Next, let us consider compositeness of hadrons in terms of quarks.
As an infrared effective theory, the constituent quark model 
has been successful for the description of the hadron spectroscopy. 
The constituent quark belongs to the fundamental representation $\bm{3}_c$ 
in the $\gr{SU}{3}$ color group, and many hadrons can be classified as the 
color-singlet ($\bm{1}_c$) bound states of some quarks and antiquarks. 
In this picture, ordinary mesons and baryons are identified as 
quark-antiquark and three-quark systems, respectively. 
However, besides the ordinary baryons and mesons, 
QCD allows the existence of other color-singlet states, 
such as glueballs, hybrids and multi-quark states, called exotic hadrons. 
Recent experiments have reported the candidates 
for these exotic states~\cite{PDG}. 
The heavy hadrons, which includes one or more heavy (anti)quarks, are also 
recent hot topics in hadron physics~\cite{PDG,Brambilla2011,Cho2011}.
For example, very recently, LHCb has reported the discovery of 
{\it two charmed pentaquarks}, P$_c^+$(4380) and P$_c^+$(4450), 
from a careful analysis of the decay product in the high-energy process, 
and this report seems to activate the multi-quark physics again \cite{LHCb2015}

In the theoretical study of these states, the diquark 
picture~\cite{Lichtenberg1967,Ida1966} has been discussed as an important 
effective degree of freedom. 
The diquark is composed of two quarks with strong correlation, 
where the one-gluon-exchange interaction between two quarks is attractive 
in the color anti-triplet $\bm{\bar{3}}_c$ 
channel~\cite{DeRujula1975,DeGrand1975}, of which color is the same 
as an anti-quark. In $\gr{SU}{3}$ flavor case, the flavor-antisymmetric 
and spin-singlet with even parity is the most attractive channel 
in diquark, which is called scalar diquark. 
If the diquark correlation is developed in a hadron, 
this scalar diquark channel would be favored. 
The diquark correlation in a hadron is discussed in various situations, such as tetra-quarks, heavy baryons and other exotic states~\cite{Jaffe2005,Close2005}.
The tetra-quark states as the bound state of the diquark/antidiquark is 
suggested in early day~\cite{Jaffe1977}, and X(3872)~\cite{Maiani2005} and 
X(1576)~\cite{Ding2006} are considered as tetra-quark states. 
Light flavor mesons as tetra-quark~\cite{Zhang2007,Lee2006,Heupel2012,Alford2000,Suganuma2007,Prelovsek2010,Wakayama2012,Alexandrou2013,Chen2007,Chen2010} and 
mixing with $q\bar{q}$ state~\cite{Sugiyama2007,Fariborz2008,Hooft2008} are 
discussed. 
There are various studies the heavy baryons focused on 
diquark~\cite{Rosner2007,Guo1999,Weng2011,Mehen2006,Selem2006}, 
e.g., single heavy quark/light diquark ($Qqq$) picture~\cite{Hernandez2008,Kim2011,Martynenko1996,GomshiNobary2007,Ebert1998}.
 The other exotic states including heavy quark(s) are studied~\cite{Jaffe2003,Lee2009,Shuryak2004,Hong2004,Karliner2003,Hyodo2013,Yasui2007}.  
The ordinary baryon properties focused on the diquarks have been also discussed~\cite{Bacchetta2008,Lichtenberg1982,Ferretti2011,Nagata2004,Fredriksson1982}. The diquark correlation is found in the lattice QCD simulation~\cite{Alexandrou2006,Babich2007,DeGrand2008,Hess1998}. It is also considered that the diquark condensation is occurred in an extremely high density system, called the color superconductivity~\cite{Alford2008}.
We note that diquark properties strongly depend on the color number $N_c$.
If we consider the two-color QCD, the diquarks compose the color singlet (baryons). The strength of correlation between two quarks is same as quark/antiquark channel, and the (diquark-)baryons correspond to the mesons. This fact is k
nown as the Pauli-G\"{u}rsey symmetry. The quark-hadron matter in two-color system is investigated~\cite{Imai2013,Ratti2004,Harada2010,Splittorff2002,Hands2007a,Kogut2001a,Strodthoff2012}.
For the $N_c=4$ case, the diquarks belong to $\bm{6}_c$ or $\bm{10}_c$. 
As an interesting fact for the case, the diquark contents must be different between baryons and tetra-quarks. 
In fact, the diquark $qq$ in an $N_c=4$ baryon $qqqq$ belongs to $\bm{6}_c$, which is self-adjoint.
On the other hand, the diquark in a tetra-quark $qq\bar{q}\bar{q}$ belongs to $\bm{10}_c$. 
From this viewpoint, the $N_c=3$ case is rather special, because the diquarks belong to the same color $\bm{\bar{3}}_c$ in both cases of baryon $qqq$ and tetra-quark $qq\bar{q}\bar{q}$.

The properties of diquarks such as the mass and size are not understood well, although the diquarks have been discussed as important object of hadron physics.
While the diquark is made by two quarks with gluonic interaction, it still strongly interacts with gluons additionally because of its non-zero color charge.
Therefore, such dressing effect of gluons for diquark should be considered in a non-perturbative way.
The dynamics of diquark and gluons may affect the structure of hadrons.
In the quark-hadron physics, the Schwinger-Dyson (SD) formalism is often used to evaluate the non-perturbative effect based on QCD~\cite{Miransky199402,Higashijima1984,Alkofer2001,Maris2003,Roberts1994,Aoki1990,Kugo1992,Fischer2006,Munczek1995,Yamanaka2013,Imai2014}. In this paper, we apply the SD formalism to scalar diquark to investigate the effective mass of scalar diquark, which reflects a non-perturbative dressing effect by gluons.
The scalar diquark is treated as an extended field like a meson in effective hadron models, and interacts with the gluons~\cite{Iida2007,Kim2011}.

For the argument of the scalar diquark, it would be important to consider its effective size. For, point scalar particles generally have large radiative corrections even in the perturbation theory \cite{Cheng1988,Peskin1995}. 
As an example, in the framework of the grand unified theory (GUT), the Higgs scalar field suffers from a large radiative correction of the GUT energy scale, and therefore severe ``fine-tuning'' is inevitably required to realize the low-lying Higgs mass of about 126GeV \cite{ATLAS2012}, which leads to the notorious hierarchy problem~\cite{Cheng1988,Peskin1995}.
The Higgs propagator with radiative correction has been investigated by setting the mass renormalization condition to reproduce 126 GeV~\cite{Maas2011,Macher2012,Hopfer2013}.
A similar large radiative correction also appears for point-like scalar-quarks, which correspond to compact scalar diquarks, in scalar lattice QCD calculations~\cite{Iida2007}. 
In fact, the point-like scalar-quark interacting with gluons acquires a large extra mass of about 1.5 GeV at the cutoff $a^{-1}\simeq 1$ GeV, where $a$ is the lattice spacing. Such a large-mass acquirement would be problematic in describing hadrons with scalar diquarks. 
However, since it is a bound-state-like object inside a hadron, 
the diquark must have an effective size. 
This effective gives a natural UV cutoff of the theory, 
and reduces the large radiative correction.
Then, we take account of the effective size and investigate 
the mass of the scalar diquark inside a hadron within the SD formalism.

This paper is organized as follows. In Sec.~\ref{sec:quark_sd}, 
we review the SD formalism for the light quark, 
as the typical fermion mass generation in QCD. 
In Sec.~\ref{sec:diquark_sd}, we investigate the SD equation for 
the scalar diquark, where a simple form factor is introduced 
for the possible size of diquark.
In Sec.~\ref{sec:num}, we present the numerical result of 
the diquark self-energy with the dependence of the bare mass and size 
of diquark, and briefly discuss the dynamical mass generation for the 
scalar diquark in the SD formalism. 
Section~\ref{sec:summ} is devoted to conclusion and discussion.

\renewcommand{\headrulewidth}{0.5pt}
\renewcommand{\footrulewidth}{0pt}

 \pagestyle{fancy}
 \fancyfoot{}
 \fancyhead{} 
 \fancyhf{}
  \fancyhead[RO]{\leavevmode \put(-90,0){\color{myaqua}S. Imai, H. Suganuma} \boxx{15}{-10}{10}{50}{15} }
 \fancyhead[LE]{\leavevmode \put(0,0){\color{myaqua}S. Imai, H. Suganuma}  \boxx{-45}{-10}{10}{50}{15} }
 \fancyfoot[C]{\leavevmode
 \put(0,0){\color{lightaqua}\circle*{34}}
 \put(0,0){\color{myaqua}\circle{34}}
 \put(-2.5,-3){\color{myaqua}\thepage}}

 \renewcommand{\headrule}{\hbox to\headwidth{\color{myaqua}\leaders\hrule height \headrulewidth\hfill}}

\sectionn{Dynamical Mass Generation of Quarks in QCD}\label{sec:quark_sd}

The chiral symmetry is a fundamental symmetry in the light-quark sector of QCD, and it is an exact global symmetry in the chiral limit. 
In the low-energy region of QCD, spontaneous chiral-symmetry breaking takes place, which generates a large effective mass of light quarks. 
Actually, in the theoretical analysis with 
the Schwinger-Dyson (SD) formalism in QCD, 
a large self-energy generation of quarks is demonstrated 
in an infrared region, which breaks the chiral symmetry 
in the physically stable vacuum \cite{Miransky199402, Higashijima1984}. 
In this section, as the standard fermionic mass generation in QCD, 
we briefly review the quark mass generation in the SD formalism 
for QCD in the Landau gauge, which is frequently used. 
This review part gives a important basis for the non-perturbative QCD physics, 
and is also useful to set up the formalism 
for the scalar diquark case in Sec.~\ref{sec:diquark_sd}. 

As a merit of the Lorentz-covariant gauge like the Landau gauge, 
the dressed quark propagator is generally described as 
$S(p^2)=iZ(p^2)(\Slash{p}-\Sigma_{q}(p^2))^{-1}$ 
with the wave function renormalization $Z(p^2)$ and 
the self-energy of quark $\Sigma_{q}(p^2)$.
The general and exact SD equation for the quark propagation is 
diagrammatically expressed in Fig.~\ref{fig:sd_q}.
In principle, the quark propagator is exactly obtained by solving 
this equation, if the exact form of the gluon propagator 
and the quark-gluon vertex are given.
Here, the kernel in the SD equation depicted in Fig.~\ref{fig:sd_q} 
is expressed by the product of 
the quark-gluon vertex $\Gamma^{\mu}_a(p,k)$ and 
the gluon dressing function $Z_g((p-k)^2)$ \cite{Iida2005}, 
\begin{align}
 \frac{g^2}{4\pi}Z_g((p-k)^2)\gamma^{\mu}T_a\Gamma^{\nu}_{b}(p,k),
\end{align}
where $T_a$ ($a=1,2,\cdots,N_c^2-1$) denotes 
the generator of the SU($N_c$) color group. 

In the most SD studies for quarks, one takes 
the rainbow-ladder approximation with 
the renormalization-group improvement of the quark-gluon vertex 
at the one-loop level. 
Note that, owing to the iterative structure of the SD equation, 
a simplified full-order treatment on the coupling $\alpha_s$ can be achieved, 
even with the use of the one-loop level vertex and so on. 
In actual, by the diagrammatical expansion, one can easily confirm the 
inclusion of infinite order of the coupling $\alpha_s$, and 
the non-perturbative effect of gluons is thus included in this formalism.
Recall that any nontrivial vacuum cannot be expressed 
by the perturbation theory.

Here, we briefly mention the treatment of 
quark confinement in the SD approach. 
In most works of the SD approach, the confinement effect is ignored, 
which seems problematic for the study of QCD.
On this point, several recent studies, both analytical works \cite{Doi2014}
and lattice QCD simulations \cite{Gongyo2012}, have suggested that 
chiral symmetry breaking and quark confinement are 
not directly correlated in QCD. 
If this is the case, even without confinement, one may be able to discuss 
chiral symmetry breaking in QCD, as is the SD approach. 

At the one-loop level of renormalization-group improvement, 
the SD kernel is approximated as 
\begin{align}
 \frac{g^2}{4\pi}Z_g((p-k)^2)\gamma^{\mu}T_a\Gamma^{\nu}_{b}(p,k)\to \alpha_s((p-k)^2)\gamma^{\mu}T_{a}\gamma^{\nu}T_{b},
\end{align}
and the Landau-gauge gluon propagator is given as
\begin{align}
 D_{\mu\nu}^{ab}(p^2)=\frac{-1}{p^2}\left(g_{\mu\nu}-\frac{p_{\mu}p_{\nu}}{p^2}\right)\delta^{ab}.
\end{align}
Then, by taking Dirac trace or the trace after multiplying $\Slash{p}$, 
the SD equation for the quark is expressed by the coupled integral equations: 
\begin{align}
 \frac{\Sigma_q(p^2)}{Z(p^2)}=&m_q+\frac{3iC_2(\bm{3})}{4\pi^3}\int d^4k\frac{\alpha_s((p-k)^2)Z(k^2)\Sigma_q(k^2)}{(k^2-\Sigma_q^2(k^2))(p-k)^2},\\
 \frac{1}{Z(p^2)}=& 1+\frac{iC_2(\bm{3})}{4\pi^3p^2}\int d^4k\frac{\alpha_s((p-k)^2)Z(k^2)}{k^2-\Sigma_q^2(k^2)}
  \left(\frac{3p\cdot k}{(p-k)^2}+\frac{2(p\cdot k)^2}{(p-k)^4}-\frac{2p^2k^2}{(p-k)^4}\right),
\end{align}
with the bare quark mass $m_q$ and 
the Casimir operator $C_2(\bm{3})=\sum_{a=1}^{8}T^aT^a=4/3$ 
in the SU(3) color case. 
\begin{figure}
\begin{center}
 \includegraphics[width=0.8\textwidth]{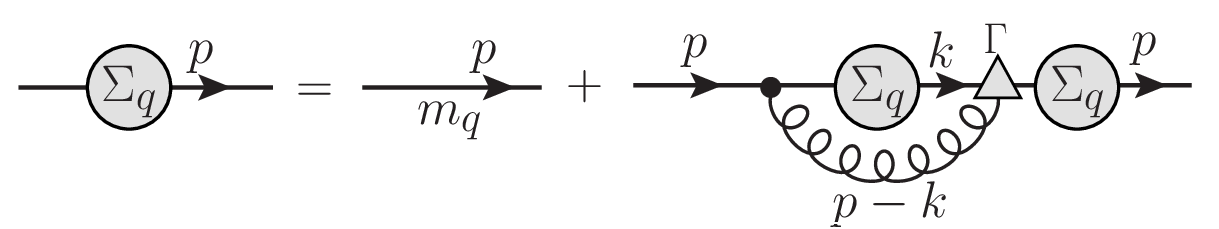}
 \end{center}
\caption{The Schwinger-Dyson equation for the quark field. 
The shaded blob denotes the self-energy of the quark $\Sigma_q(p^2)$, 
the black dot the bare quark-gluon vertex, the shaded triangle the dressed 
vertex $\Gamma^{\mu}_a(p,k)$, the solid line the quark propagator 
and the curly line the gluon propagator.}
\label{fig:sd_q}
\end{figure}

We use one-loop level renormalization-group-improved coupling 
in the case of $N_c=3$ and $N_f=3$,
\begin{align}
 \alpha_s(p_E^2)=&\frac{g^2(p_E^2)}{4\pi}=\frac{12\pi}{11N_c-2N_f}
\begin{cases}
 \frac{1}{\ln(p_E^2/\Lambda_{\rm QCD}^2)} 
\quad (p_E^2 \geq p_{\rm IR}^2)\\
 \frac{1}{\ln(p_{\rm IR}^2/\Lambda_{\rm QCD}^2)} 
 \quad (p_E^2 \leq p_{\rm IR}^2)
\end{cases},\label{eq:coupling}
\end{align}
with an infrared regularization of a simple cut at $p_{\rm IR}\simeq 640$ MeV which leads to $\ln(p_{\rm IR}^2/\Lambda_{\rm QCD}^2)=1/2$, and the QCD scale parameter $\Lambda_{\rm QCD}=500$ MeV~\cite{Higashijima1984,Aoki1990,Yamanaka2013}.  The subscript $E$, such as $p_E$, denotes the value in Euclidean space. 
The infrared regularization has been introduced to avoid the divergent pole at $p=\Lambda_{\rm QCD}$.
 The behavior of the coupling is shown in Fig.~\ref{fig:coupling} in the Euclidean space. All the figures for the numerical results will be in the Euclidean space.

\begin{figure}
\begin{center}
 \includegraphics[width=0.4\textwidth]{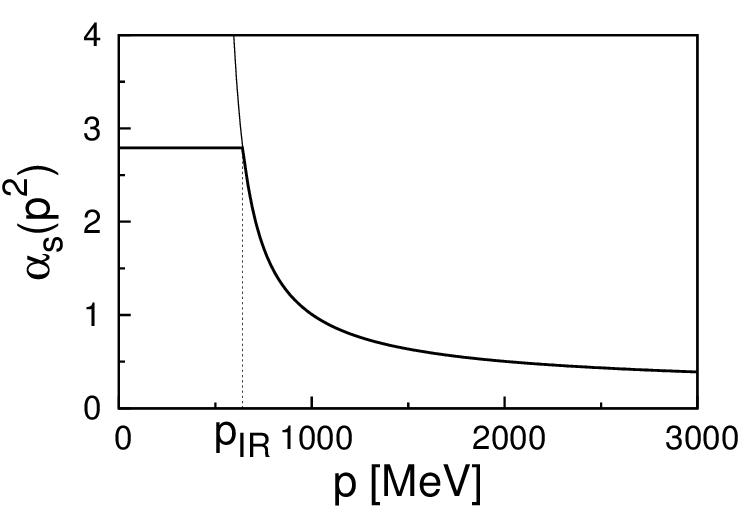}
 \end{center}
 \caption{The behavior of the running coupling of our model $\alpha_s(p^2)$ as a function of the momentum $p$ in the Euclidean space. The thin line is the one-loop renormalization group improved running coupling. We introduce a simple cut at $p_{\rm IR}$ as an infrared regularization.}\label{fig:coupling}
\end{figure}

The Higashijima-Miransky approximation is to take the larger value of 
the argument (Euclidean momenta) in the coupling as 
$\alpha_s((p_E-k_E)^2)\approx \alpha_s(\max(p_E^2,k_E^2))$, 
and this approximation is also frequently used in the SD approach for quarks, 
because $Z(p_E^2)=1$ is analytically obtained in the Landau gauge and 
the computation becomes quite simplified for 
the quark self-energy $\Sigma_q(p_E^2)$: 
 \begin{align}
  \Sigma_q(p^2_E)
  =m_q+\frac{2\alpha_s(p_E^2)}{\pi p_E^2}\int_{0}^{p_E}dk_E\frac{k_E^3\Sigma_q(k_E^2)}{k_E^2+\Sigma_q^2(p_E^2)}
  +\frac{2}{\pi}\int_{p_E}^{\Lambda_{\rm UV}}dk_E\frac{k_E\alpha_s(k_E^2)\Sigma_q(k_E^2)}{k_E^2+\Sigma_q^2(k_E^2)},\label{eq:sd_q}
 \end{align}
where the Wick rotation has been taken. 
 (For the detail, see, e.g., Appendix in Ref.~\cite{Aoki1990}.) 
The result of the SD equation is shown in Fig.~\ref{fig:self_q} in the chiral limit $m_q=0$. There is a small cusp structure at $p_{\rm IR}$ due to the coupling behavior Eq.~\eqref{eq:coupling}.
The ultraviolet cutoff $\Lambda_{\rm UV}$ is taken as 5 GeV. The self-energy $\Sigma_q(p_E^2)$ is unchanged even the cutoff is taken 10 GeV. 
The quark mass is large at the infrared region and monotonously goes to zero with the momentum, which reflects spontaneous chiral-symmetry breaking~\cite{Nambu1961,Miransky199402,Higashijima1984,Hatsuda1994}.
\begin{figure}
\begin{center}
 \includegraphics[width=0.4\textwidth]{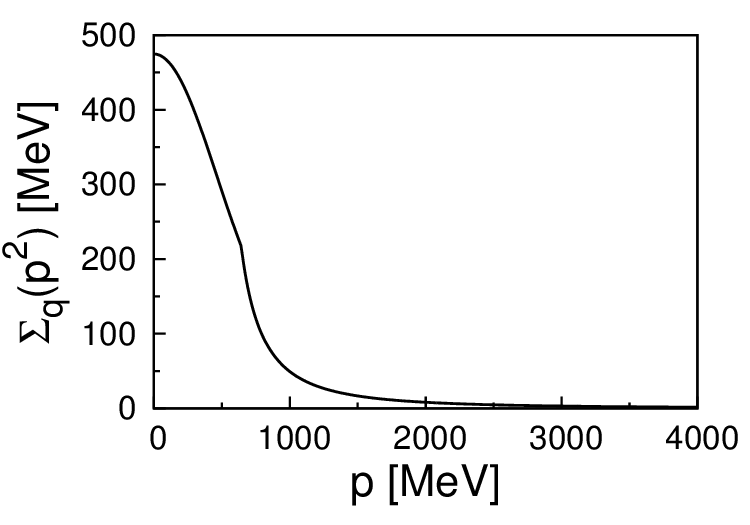}
 \end{center}
\caption{The quark self-energy $\Sigma_q(p^2)$ as a function of the momentum $p$ in the chiral limit. The self-energy is large in the low momentum region and goes to zero monotonously with the momentum.}\label{fig:self_q}
\end{figure}

The scale parameter $\Lambda_{\rm QCD}$ is chosen to reproduce chiral properties for quarks in the SD formalism with the Higashijima-Miransky approximation in the Landau gauge, while the ordinary QCD scale parameter is around $\Lambda_{\rm QCD}\sim 200-300$ MeV.
The self-energy leads to the pion decay constant with the Pagels-Stokar approximation~\cite{Pagels:1979hd}:
\begin{align}
 f_{\pi}^2=\frac{N_c}{2\pi^2}\int_{0}^{\infty}dk_E
\frac{k_E^3\Sigma_q(k_E^2)}{[k_E^2+\Sigma_q^2(k_E^2)]^2}
\left(\Sigma_q(k_E^2)-\frac{k_E}{4}\frac{d}{dk_E}\Sigma_q(k_E^2)\right),\label{eq:fpi}
\end{align}
and the (unrenormalized) chiral condensate:
\begin{align}
\vev{\bar{q}q}_{\Lambda_{\rm UV}}=-\frac{N_c}{2\pi^2}\int_{0}^{\Lambda_{\rm UV}}dk_E\frac{k_E^3\Sigma_q(k_E^2)}{k_E^2+\Sigma_q^2(k_E^2)}.\label{eq:chiral}
\end{align}
Since the pion decay constant is a physical value, its renormalization is not required and it does not depend on the ultraviolet cutoff $\Lambda_{\rm UV}$. 
Hence, the upper limit of the integration has been taken as $\Lambda_{\rm UV}\to\infty$.
On the other hand, the chiral condensate depends on the renormalization point. 
We adopt a standard renormalization point $\mu=2$ GeV~\cite{PDG}, and consider the chiral condensate $\vev{\bar{q}q}_{\mu=2{\rm GeV}}$ according to the renormalization-group formula~\cite{Aoki1990,Kugo1992,Yamanaka2013}:
\begin{align}
\vev{\bar{q}q}_{\mu=2{\rm GeV}}=&\left(\frac{\alpha_s(\Lambda^2)}{\alpha_s(\mu^2)}\right)^{\frac{3C_{2}(N_c)}{16\pi^2\beta_0}}\vev{\bar{q}q}_{\Lambda_{\rm UV}},\label{eq:rchiral}
\end{align}
with $\frac{3C_{2}(N_c)}{16\pi^2\beta_0}=4/9$ and $\beta_{0}=\frac{11N_{c}-2N_{f}}{48\pi^2}$ corresponding to the lowest coefficient of the $\beta$ function of the renormalization group.
Taking the scale parameter $\Lambda_{\rm QCD}$ as 500 MeV and the ultraviolet cutoff $\Lambda_{\rm UV}$ as 5 GeV, the pion decay constant and the chiral condensate are fixed as $f_{\pi}\simeq 90$ MeV and $-\vev{\bar{q}q}_{\mu=2{\rm GeV}}^{1/3}\simeq 242$ MeV, respectively. 
We have numerically checked that they are stable against the variation of the ultraviolet cutoff $\Lambda_{\rm UV}$.
The SD formalism with the approximations in the Landau gauge reproduces these chiral properties well.

\sectionn{The Schwinger-Dyson Equation for the Scalar Diquark}\label{sec:diquark_sd}

In this section, we investigate the scalar diquark, 
i.e., an extended colored scalar object, 
and its mass generation, using the Schwinger-Dyson (SD) formalism. 

Diquark is a bound-state-like object of two quarks and decomposed into color anti-triplet $\bm{\bar{3}}_c$ and sextet $\bm{6}_c$ and flavor anti-triplet $\bm{\bar{3}}_f$ and sextet $\bm{6}_f$ in SU(3) flavor case. The most attractive channel for diquark is the color and flavor anti-triplet $\bm{\bar{3}}_{c,f}$ and spin singlet with even parity $0^{+}$ by one gluon exchange~\cite{DeRujula1975,DeGrand1975} and by instanton interactions~\cite{Hooft1976,Schaefer1998}, which is called scalar diquark.
If the diquark correlation is developed in a hadron such as a heavy baryon ($Qqq$), this scalar diquark channel would be favored.
We consider the scalar diquark as an effective degree of freedom with a peculiar size, assuming it to be an extended scalar field $\phi(x)$~\cite{Iida2007,Kim2011} like a meson in the effective hadron models. 
The scalar diquark is composed of two quarks with the gluonic interaction, and still affected by non-perturbative gluonic effects since it has non-zero color charge as shown Fig.~\ref{fig:diquark}. 
The dynamics of the scalar diquark field $\phi$ is expected to be described 
by the gauge-invariant scalar-QCD-type Lagrangian:
\begin{align}
 \mcl{L}=[(\partial^\mu+igA^{\mu}_{a}T^{a})\phi]^\dagger[(\partial_{\mu}+igA_{\mu b}T^{b})\phi]-m_{\phi}^2\phi^\dagger\phi,\label{scalarqcd}
\end{align}
where the bare diquark mass $m_{\phi}$ and the gauge field $A^{\mu}_{a}$ (gluon) with the generator $T^a$ have been introduced. We note that the scalar diquark has the 4-point interaction term 
of $|\phi|^2 A^2$ type, which is different from the quark.
In general, such gauged scalar fields accompany the 4-point interaction~\cite{Cheng1988,Maas2011,Macher2012,Hopfer2013,Itzykson2006}.
\begin{figure}
\begin{center}
\subfigure[Gluonic interaction between two quarks]
{\includegraphics[width=0.4\textwidth]{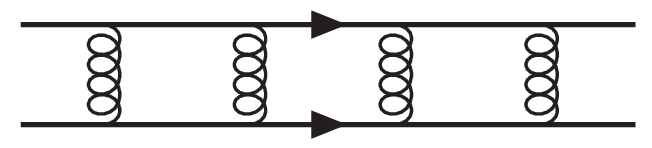}\label{fig:inter}}
\hspace{1.5cm}
\subfigure[Gluonic dressing for a diquark]
 {\includegraphics[width=0.4\textwidth]{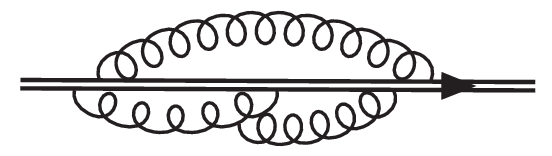}\label{fig:dressing}}
 \end{center}
\caption{The two types of gluonic interaction for a diquark: \subref{fig:inter} inter-two-quarks gluonic interaction to form a diquark and \subref{fig:dressing} gluonic dressing for the diquark due to its non-zero color charge. The single line denotes a quark, the double line a diquark and the curly line a gluon.}\label{fig:diquark}
\end{figure}

Since the diquark is a bound-state-like object confined in a hadron, 
it must have an effective size and its size should be 
smaller than the hadron. In order to include the size effect of diquark, 
we introduce a simple ``form factor'' in the four-dimensional 
Euclidean space as
\begin{align}
 f_{\Lambda}(p_E^2)=\left(\frac{\Lambda^2}{p_E^2+\Lambda^2}\right)^{\nu},\label{eq:form}
\end{align}
where the momentum cutoff $\Lambda$ corresponds to the inverse of the diquark size $R$. 
In this paper, we set $R\equiv\Lambda^{-1}$.
Since the radiative correction for the scalar particle is generally large, this form factor has also a role of the convergence factor.
As for the form factor $f_{\Lambda}(p_E^2)$, it has the roles of introducing an effective size and convergence of the SD equation, so one can use arbitrary function such as the step function $\theta(\Lambda^2-p_E^2)$, the exponential function $\exp(-p_E^2/\Lambda^2)$ and so on.
In this study, we take Eq.~\eqref{eq:form} with $\nu=2$ to simple analysis and the convergence of the SD equation.
The size effect of the diquark can be included in the vertex as $\alpha_s(p^2)\to\alpha_s(p^2)f_{\Lambda}(p^2)$.

While the scalar QCD Lagrangian \eqref{scalarqcd} is renormalizable, 
this theory is an effective cutoff theory with an UV cutoff parameter 
$\Lambda$, which corresponds to the inverse size of the scalar diquark. 
Here, the scalar diquark cannot be observed as an isolated object, 
and has no characteristic symmetry, such as the chiral symmetry, 
so that it is difficult to set the renormalization condition. 
Instead, we introduce an effective size $R=\Lambda^{-1}$ of the diquark, 
which leads to a natural UV cutoff in the theory. 
As we will see later, the effective size of diquark will play 
an important role for the convergent of loop integrations, 
and therefore we will not take the limit of $\Lambda\to \infty$ ($R\to 0$).
In fact, the extended diquark is treated as the effective degrees of 
freedom appearing in the QCD system of quarks and gluons, 
and hence, also for the scalar diquark, 
we basically use the same framework as 
the single quark case, presented in the previous section. 
For instance, we will use the same running coupling $\alpha_s(p_E^2)$ 
in Eq.(\ref{eq:coupling}) for the argument of diquarks. 

We now describe the SD equation for the scalar diquark, 
as shown in Fig.~\ref{fig:sd_d}. 
\begin{figure}
\begin{center}
 \includegraphics[width=0.9\textwidth]{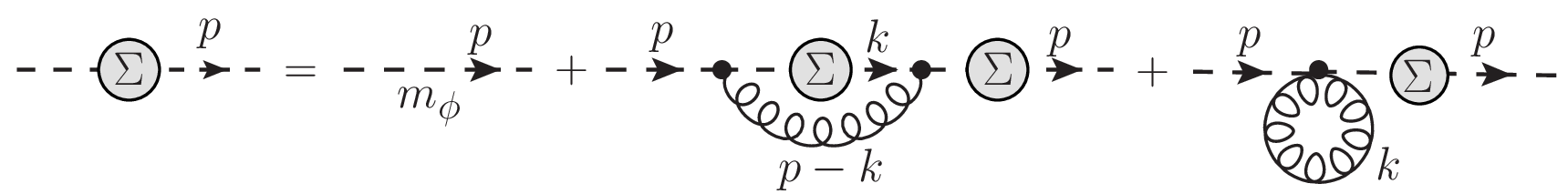}
 \end{center}
 \caption{The Schwinger-Dyson equation for the scalar diquark. 
The shaded blob is the self-energy $\Sigma(p^2)$, the dashed line denotes the scalar diquark propagator and the curly line the gluon propagator. 
The last term arisig from 4-point interaction is the peculiar term in 
gauged scalar theories, and it does not appear 
in the single quark case in QCD.}\label{fig:sd_d}
\end{figure}
For the self-energy diagram, we include the first order of the coupling $\alpha_s$ at the one-loop level, like the improved ladder QCD~\cite{Miransky199402,Higashijima1984,Aoki1990,Kugo1992}. 
Note however that, due to the iterative calculation, 
this formalism includes infinite order of 
the coupling $\alpha_s$ and describes non-perturbative effects. 
It is also notable that the same form of the running coupling for 
the quark/gluon coupling can be used even for 
the scalar diquark/gluon~\cite{Marinho2007,Salam1978}. 
(In particular, in the heavy mass limit of colored particles, 
the QCD interaction depends only on their color.) 
Since the scalar diquark corresponds to an antiquark 
in terms of the color representation, 
we may use the same form of the running coupling 
even for the scalar diquark case. 
Then, the SD equation for the scalar diquark is 
diagrammatically expressed as Fig.~\ref{fig:sd_d} and is written by
 \begin{align}
 \Sigma^2(p_E^2) = m_{\phi}^2&+\frac{3C_{2}(\bm{3})}{2\pi^3}\int_{0}^{\Lambda_{\rm UV}} d^4k_E \frac{\alpha_s(k_E^2)f_{\Lambda}(k_E^2)}{k_E^2}\nn
  &-\frac{C_{2}(\bm{3})}{\pi^3}\int_{0}^{\Lambda_{\rm UV}} d^4k_E\frac{\alpha_s((p_E-k_E)^2)f_{\Lambda}((p_E-k_E)^2)}{k_E^2+\Sigma^2(k_E^2)}
  \frac{p_E^2k_E^2-(p_E\cdot k_E)^2}{(p_E-k_E)^4}.
\label{eq:sd_d}
 \end{align}
In the right-hand side of Eq.(\ref{eq:sd_d}), 
the second term arises from the 4-point vertex and 
the third term is lead from the 3-point vertex, 
as shown in Fig.~\ref{fig:sd_d}.
Here, we do not consider the wave functional renormalization, 
as is often assumed for the quark field in the Landau gauge. 
Similarly in the single quark case, 
we adopt the Higashijima-Miransky approximation 
$\alpha_s((p_E-k_E)^2)\approx \alpha_s(\max(p_E^2,k_E^2))$ 
for the 3-point vertex, and finally obtain the SD equation 
for the self-energy $\Sigma^2(p_E^2)$ of the scalar diquark: 
 \begin{align}
 \Sigma^2(p_E^2) 
 = m_{\phi}^2&+\frac{4}{\pi}\int_{0}^{\Lambda_{\rm UV}} dk_E k_E \alpha_s(k_E^2)f_{\Lambda}(k_E^2) \nn
  &-\frac{2\alpha_{s}(p_E^2)f_{\Lambda}(p_E^2)}{\pi p_E^2}\int_{0}^{p_E}dk_E\frac{k_E^5}{k_E^2+\Sigma^2(k_E^2)}
  -\frac{2p_E^2}{\pi}\int_{p_E}^{{\Lambda}_{\rm UV}}dk_E\frac{\alpha_s(k_E^2)f_{\Lambda}(k_E^2)k_E}{k_E^2+\Sigma^2(k_E^2)}.\label{eq:sd_d2}
 \end{align}

\sectionn{Numerical Results and Discussion}\label{sec:num}
\subsectionn{The Parameter Setting}

The bare mass $m_{\phi}$ and cutoff $\Lambda$ (inverse of the size $R$) 
are free parameters of the diquark theory. 
In this subsection, we consider the possible range of 
these parameters from the physical viewpoint. 

The diquark is originally made of two consistent quarks, and 
the color-Coulomb interaction is one of the main attractive forces. 
We here estimate the color-Coulomb interaction between the two massive quarks from 
the three-quark (3Q) potential \cite{Takahashi01}, 
or generally from the mult-quark potential such as 
4Q(${\rm QQ\bar Q\bar Q}$) 
and 5Q(${\rm 4Q \bar Q}$) potentials \cite{Okiharu05}.
In SU(3) lattice QCD, the 3Q potential among the three quarks 
located at ${\bf r}_i (i=1, 2, 3)$ is well reproduced by 
\begin{align} 
V_{\rm 3Q} = -\sum_{i<j} \frac{A_{\rm 3Q}}{|{\bf r}_i-{\bf r}_j|}
+\sigma  L_{\rm min},
\end{align}
with the color-Coulomb coefficient 
$A_{\rm 3Q}\simeq A_{\rm Q \bar{Q}}/2 \simeq 0.12(1)$, 
the string tension $\sigma \simeq 0.89{\rm GeV/fm}$ and 
the minimal flux-tube length $L_{\rm min}$ \cite{Takahashi01}. 
Since the color-Coulomb potential energy between two quarks 
is $A_{\rm 3Q}/R$ for the inter-quark distance $R$, 
the potential energy is estimated as 
$A_{\rm 3Q}/R\simeq 24-80{\rm MeV}$ for 
the typical range of $R=0.3-1{\rm fm}$, 
and its value is not so large in comparison with the two-quark mass of about 600MeV. 
[Note also that similar estimation also leads to 
a small value of the diquark-diquark interaction, 
which gives a reason of the absence of $(\phi^\dagger \phi)^2$ 
in the diquark Lagrangian (\ref{scalarqcd}).]
The same result can be obtained from 
the multi-quark potential \cite{Okiharu05}, because 
the color-Coulomb coefficient is the same for two quarks in the diquark, 
i.e., $A_{n{\rm Q}}\simeq A_{\rm Q \bar{Q}}/2 \simeq 0.12(1)$ for $n$=3,4,5.
Therefore, the bare mass of diquark is expected to be 
simply considered as the twice of the quark mass. 

In this paper, we consider two cases of the bare diquark mass. 
One is twice of constituent quark mass, i.e., $m_{\phi}=600$ MeV.
The other is twice of the running quark self-energy, i.e., $m_{\phi}(p_E^2)=2\Sigma_q(p_E^2)$, where $\Sigma_q(p_E^2)$ is determined by the SD equation for single quark Eq.~\eqref{eq:sd_q}. This means that the diquark is constructed by the two dressing quarks. The constant bare mass case is based on the constituent quark model like picture and the running bare mass case is the SD formalism with omitting the effect of the gluonic attraction force between two quarks. The diquark should be dressed by gluon furthermore because of its non-zero color charge.

The cutoff $\Lambda$ corresponds to the diquark size in a hadron, $R$, i.e., $\Lambda\equiv R^{-1}$, so the diquark should be smaller than the hadron. We also consider two cases of the size. One is the typical size of a baryon, $R=1$ fm, i.e., $\Lambda=200$ MeV, which gives the upper limit of the size (the lower limit of the cutoff). The diquark covers the baryon in this case.  The second is the typical size of a constituent quark, $R\simeq 0.3$ fm, i.e., $\Lambda=600$ MeV, which gives the lower limit of the size (the upper limit of the cutoff).

\subsectionn{The Constant Bare Mass Case}\label{sec:constant}
We first show in Fig.~\ref{fig:cm} the case of the constant bare mass $m_{\phi}=600$ MeV with dependence on the cutoff $\Lambda$.
The diquark self-energy $\Sigma(p^2)$ is always larger than the bare mass $m_{\phi}$ and almost constant except for a small bump structure 
in an infrared region. 
The value of the self-energy is strongly depends on the cutoff $\Lambda$, e.g., the ``compact diquark'' with $R\simeq 0.3$ fm has a large mass.
\begin{figure}
\vspace{-0.3cm}
\begin{center}
  \subfigure[$\Lambda=200$ MeV ($R=1$ fm)]{\includegraphics[width=0.37\textwidth]{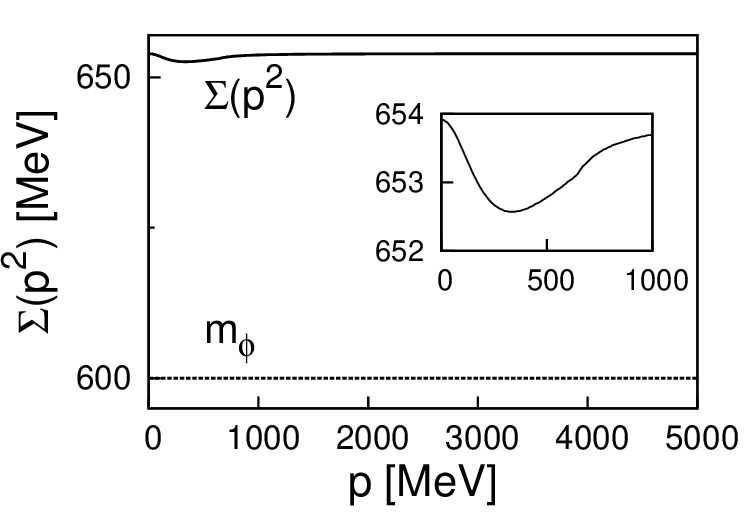}\label{fig:c2}}
\hspace{0.5cm}
 \subfigure[$\Lambda=600$ MeV ($R\simeq 0.3$ fm)]{\includegraphics[width=0.37\textwidth]{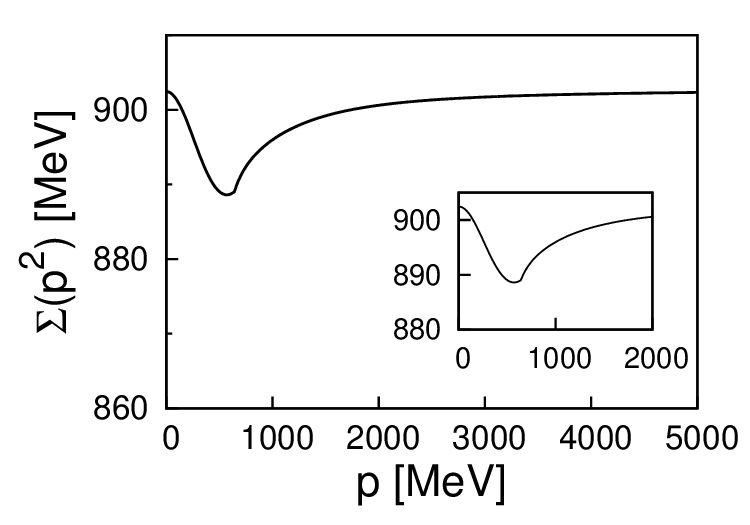}\label{fig:c6}}
 \end{center}
\vspace{-0.1cm}
\caption{The scalar diquark self-energy $\Sigma(p^2)$ as a function of the momentum $p$ in the constant bare mass case of $m_{\phi}=600$ MeV with \subref{fig:c2} $\Lambda=200$ MeV, i.e., $R=1$ fm and \subref{fig:c6} $\Lambda=600$ MeV, i.e., $R\simeq 0.3$ fm. In both cases, there appears a small bump structure, which is displayed in the small window. In the left figure, the original bare mass $m_{\phi}$ is plotted for comparison.}\label{fig:cm}
\vspace{-0.1cm}
\end{figure}

The scalar QCD includes both 3-point and 4-point interactions, and the existence of 4-point interaction is diagrammatically different from the ordinary QCD.
To see the role of each interaction, we consider the calculation of the artificial removal of 3-point interaction and 4-point interaction, respectively. 
In fact, we investigate the two cases: 
\subref{fig:o3} removal of 4-point interaction and \subref{fig:o4} removal of 3-point interaction.
The result is shown in Fig.~\ref{fig:only} in the case of $\Lambda=200$ MeV. 
The bump structure appears in the case without the 4-point interaction term as shown in Fig.~\ref{fig:o3}.
Although the diagrammatic expression of the SD equation for the scalar diquark without 4-point interaction term is analogous to the quark SD equation, the behavior is completely different from the quark case. The diquark self-energy $\Sigma(p_E^2)$ starts from the bare mass $m_{\phi}=600$ MeV at zero momentum, then decreases at low momentum and rises up to the original value 600 MeV. On the other hand, the quark self-energy $\Sigma_q(p_E^2)$ starts from a large value and goes to zero monotonously with the momentum. The SD equation without 3-point interaction just rises the self-energy and keeps constant. 
The strong dependence of the cutoff $\Lambda$ (or the size $R$) mainly comes from the 4-point interaction term.
\begin{figure}
\vspace{-0.3cm}
\begin{center}
\subfigure[Without 4-point interaction term]{\includegraphics[width=0.37\textwidth]{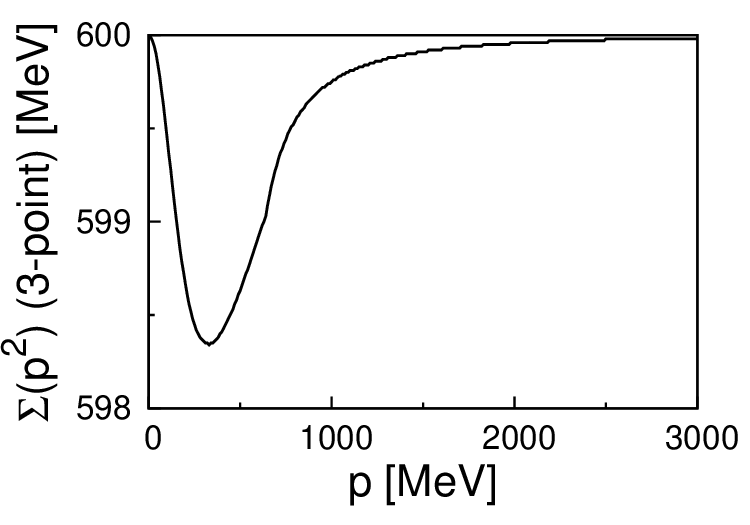}\label{fig:o3}}
\hspace{0.75cm}
 \subfigure[Without 3-point interaction term]{\includegraphics[width=0.37\textwidth]{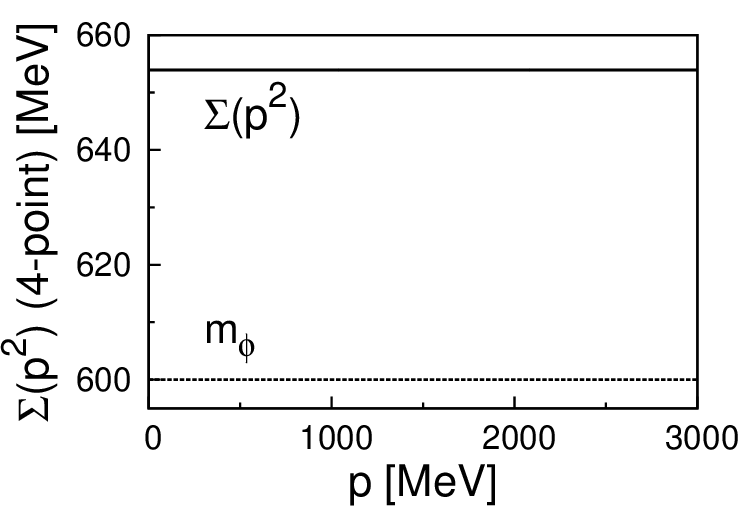}\label{fig:o4}}
 \end{center}
\vspace{-0.1cm}
\caption{The self-energy $\Sigma(p^2)$ in the case of \subref{fig:o3} without 4-point interaction and \subref{fig:o4} without 3-point interaction as the function of the momentum $p$. Here, $\Lambda=200$ MeV is taken. In the right figure, the original bare mass $m_{\phi}$ is plotted for comparison.}\label{fig:only}
\vspace{-0.1cm}
\end{figure}

\subsectionn{The Running Bare Mass Case}
We show in Fig.~\ref{fig:rm} the case of the running bare mass $m_{\phi}(p_E^2)=2\Sigma_q(p_E^2)$ with dependence on the cutoff $\Lambda$. The diquark self-energy $\Sigma(p_E^2)$ also strongly depends on the cutoff $\Lambda$.
In the low-momentum region, the behavior of $\Sigma(p_E^2)$ reflects the running property of the bare mass, especially in the $\Lambda=200$ MeV case, the gluonic effect seems to be small, because of $\Sigma(p_E^2)\approx 2\Sigma_q(p_E^2)$.
In the high-momentum region, the diquark self-energy keeps a large value, while the bare mass $m_{\phi}(p_E^2)$ goes to zero. This suggests the mass generation of the scalar diquark by gluonic radiative correction.
\begin{figure}
\vspace{-0.3cm}
\begin{center}
  \subfigure[$\Lambda=200$ MeV ($R=1$ fm)]{\includegraphics[width=0.4\textwidth]{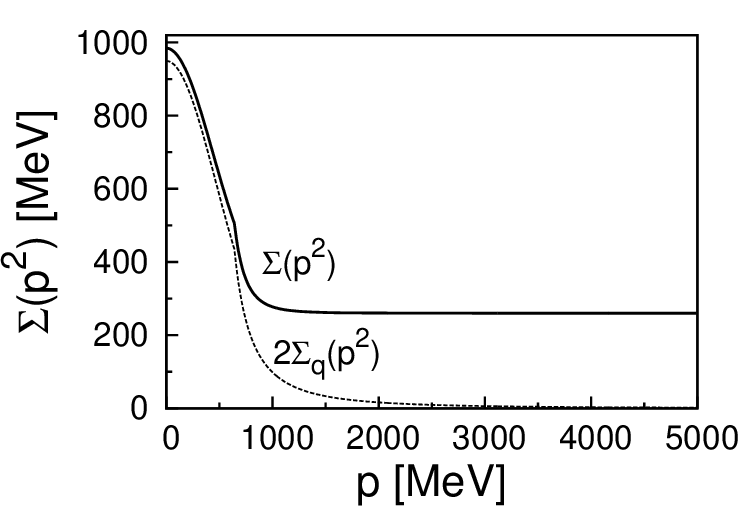}\label{fig:q2}}
\hspace{0.5cm}
 \subfigure[$\Lambda=600$ MeV ($R\simeq 0.3$ fm)]{\includegraphics[width=0.4\textwidth]{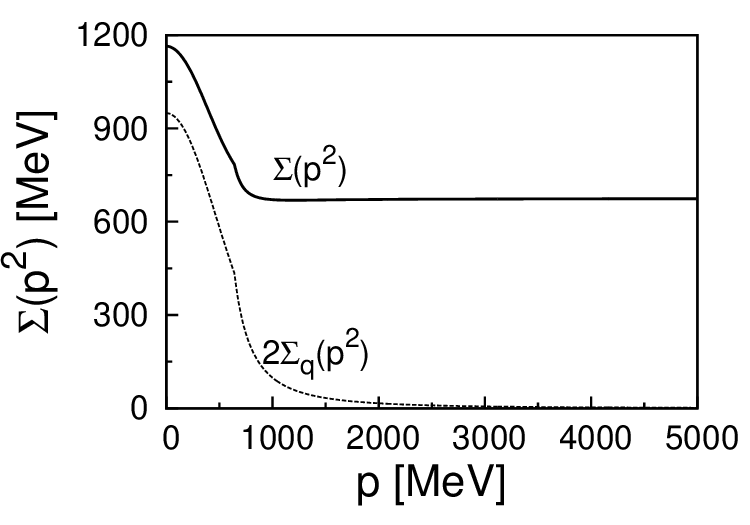}\label{fig:q6}}
 \end{center}
\vspace{-0.1cm}
  \caption{The scalar diquark self-energy $\Sigma(p^2)$ as a function of the momentum $p$ in the running bare mass case with \subref{fig:q2} $\Lambda=200$ MeV and \subref{fig:q6} $\Lambda=600$ MeV. The bare mass $m_{\phi}(p)=2\Sigma_q(p)$ is also plotted with the dotted line for comparison.}\label{fig:rm}
\vspace{-0.1cm}
\end{figure}

\subsectionn{Discussion on the Scalar Diquark Property}

In this subsection, we discuss the mass and the size of 
the scalar diquark, with comparing to the chiral quark. 
One of the most important properties of single quark SD equation 
\eqref{eq:sd_q} is the existence of the trivial solution $\Sigma_{q}=0$ 
in the chiral limit $m_{q}\to0$. 
In fact, the quark mass remains to be zero due to the chiral symmetry 
in the perturbative treatment, 
and the quark mass generation, i.e., chiral symmetry breaking, 
is realized by the non-perturbative gluonic interaction 
\cite{Miransky199402,Higashijima1984}. 
Such arguments can be done even in the limit of $\Lambda_{\rm UV} \to \infty$, 
which is consistent with the point quark as an elementary particle. 

On the other hand, the SD equation \eqref{eq:sd_d} for scalar diquark 
has no trivial solution and is a highly non-linear equation, 
even in the zero bare mass limit $m_{\phi}\to0$. 
For example, the 4-point interaction term gives a strong dependence of 
the UV cutoff $\Lambda$. 
This is similar to the framework of GUT, where the Higgs scalar field 
suffers from a large radiative correction of the GUT energy scale.

Actually, the scalar diquark self-energy $\Sigma(p^2)$ 
strongly depends on the diquark size $R\equiv\Lambda^{-1}$ 
in both cases of the bare mass. 
In an extreme case of the point-like limit $R\to0$, i.e., $\Lambda\to\infty$, 
the diquark effective mass diverges.
This suggests that the simple treatment of point-like diquarks is somehow 
dangerous in hadron models and the diquark must have an effective size.

As a quantitative argument, our calculations show that the ``compact diquark'' 
with $R\simeq 0.3$ fm has a large effective mass in both cases, and  
does not seem to be acceptable in effective models for hadrons. 
In fact, the appropriate diquark is not so compact as $R\simeq 0.3$ fm 
but is fairly extended as $R\sim 1$fm.

\subsectionn{Mass Generation for Colored Scalar Particle}\label{sec:mass}
Finally, we consider the zero bare-mass case of diquark, $m_{\phi}\equiv 0$.
Even for a finite mass of quark, the bare mass of diquark can be zero, 
if the attraction between two quarks extremely strong.
The result is shown in Fig.~\ref{fig:zero} for the two cases: \subref{fig:m2} $\Lambda=200$ MeV and \subref{fig:m6} $\Lambda=600$ MeV on the cutoff. 
The self-energy $\Sigma(p_E^2)$ is always finite and takes a large value 
even for $m_{\phi}\equiv 0$.
The mass generation mechanism in QCD is usually considered in the context of 
spontaneous chiral-symmetry breaking.
On the other hand, our scalar diquark theory is composed of an effective scalar diquark field $\phi(x)$ and does not have the chiral symmetry explicitly, 
although the original diquark is constructed by two chiral quarks. 
Nevertheless, the effective mass of diquark emerges by the non-perturbative 
gluonic effect. 
In fact, the mechanism of dynamical mass generation seems 
to work in the scalar diquark theory, even without chiral symmetry breaking.
If we take $\Lambda=1$ GeV, the diquark self-energy is $\Sigma\sim 950$ MeV. 
This result seems to be consistent with the lattice QCD result 
on the colored scalar particle~\cite{Iida2007}.

\begin{figure}
\vspace{-0.3cm}
\begin{center}
  \subfigure[$\Lambda=200$ MeV ($R=1$ fm)]{\includegraphics[width=0.4\textwidth]{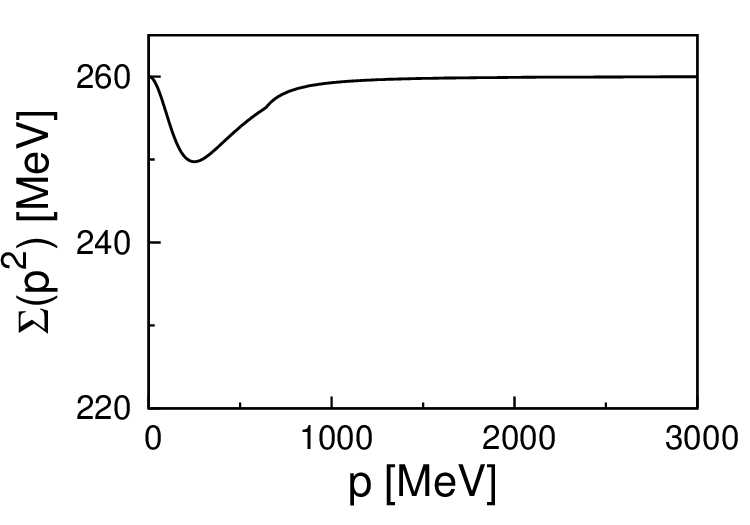}\label{fig:m2}}
\hspace{0.5cm}
 \subfigure[$\Lambda=600$ MeV ($R\simeq 0.3$ fm)]{\includegraphics[width=0.4\textwidth]{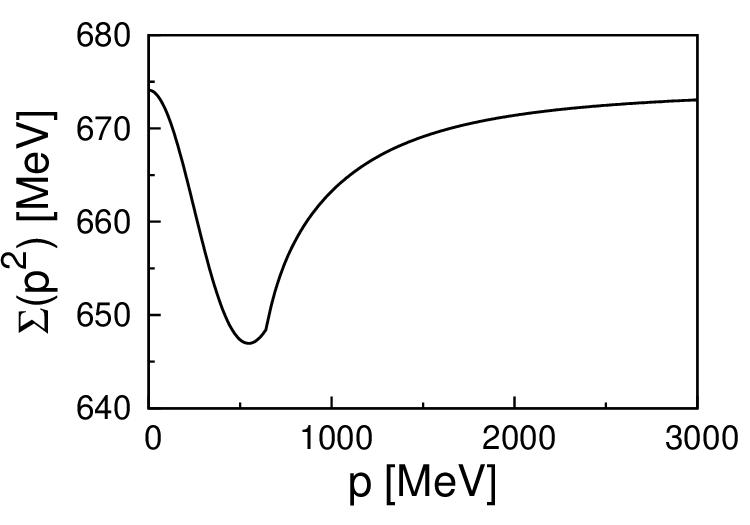}\label{fig:m6}}
 \end{center}
\vspace{-0.1cm}
  \caption{The scalar diquark self-energy $\Sigma(p^2)$ as a function of the momentum $p$ in the massless case of $m_{\phi}=0$. The self-energy $\Sigma(p^2)$ is finite in both cases.}\label{fig:zero}
\vspace{-0.1cm}
\end{figure}

\sectionn{Conclusion and Discussion}\label{sec:summ}

We have studied various mass generation of 
colored particles and gluonic dressing effect 
in a non-perturbative manner, 
using the Schwinger-Dyson (SD) formalism in QCD. 
First, we have briefly reviewed 
dynamical quark-mass generation in QCD in the SD approach 
as a typical fermion-mass generation via spontaneous chiral-symmetry breaking. 
Second, using the SD formalism for scalar QCD, 
we have investigated the scalar diquark, 
a bound-state-like object of two quarks, 
and its mass generation, which is clearly non-chiral-origin. 
Considering the possible size of the diquark inside a hadron, 
the effect of diquark size $R$ is introduced as 
a cutoff parameter $\Lambda=R^{-1}$ in the form factor, 
as is used in effective theories.
 
The basic technology of scalar SD formalism is imported from the single quark case, such as the running coupling, the approximations and so on.
Since the diquark is located in and construct of a hadron, the size should be smaller than the hadron ($R\sim 1$ fm) and larger than the constituent quark ($R\sim 0.3$ fm).
The size (cutoff) dependence of self-energy have been investigated.
 We have considered the two cases of the constant bare mass $m_{\phi}=600$ MeV and the running bare mass $m_{\phi}(p_E^2)=2\Sigma_q(p_E^2)$. The diquark self-energy strongly depends on the size $R=\Lambda^{-1}$ in both cases, especially the small diquark ($R\simeq0.3$ fm) has a large effective mass by the gluonic dressing effect.

We find that the effective diquark mass is finite and large even for the zero bare-mass case, and the value strongly depends on the size $R$, which is an example of dynamical mass generation by the gluonic effect, without chiral symmetry breaking.
The mass difference between current and constituent charm quark mass and the large glueball mass are also examples of this type of mass generation.
In this sense, spontaneous chiral-symmetry breaking may be a special case of massless (or small mass) fermion. As was conjectured in Ref.\cite{Iida2007}, 
it would be a general property of strong interacting theory that all colored particles acquire a large effective mass by the dressing effect, 
as shown in Fig.~\ref{fig:mass}.
\begin{figure}
\begin{center}
 \includegraphics[width=.6\textwidth]{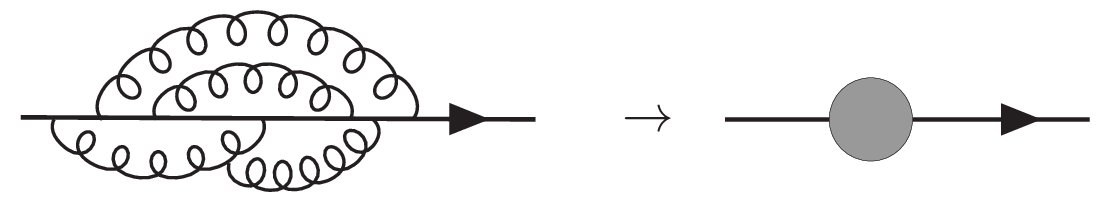}
 \end{center}
\caption{The schematic picture for dynamical mass generation of the colored particle. The colored particle (solid line) interacting with the gluons (curly line). The effective mass emerges by the non-perturbative interaction even without the chiral symmetry.}\label{fig:mass}
\end{figure}

In this study, we have mainly investigated the diquark properties, and have not calculated physical quantities. It is however desired to describe the color-singlet states such as heavy baryon $Qqq$ based on the scalar theory. 
One of description of diquark based on QCD is the Bethe-Salpeter (BS) formalism for two quarks~\cite{Bender1996,Cahill1987,Wang2007,Bloch1999}. However, the treatment of the scalar diquark as an explicit degree of freedom $\phi(x)$ is a good approximation for the structure of the heavy baryons. 
The constituent scalar-quark(diquark)/quark picture in the scalar lattice QCD~\cite{Iida2007} and the structure of $\Lambda_{h}$ ($h=s,c,b$ quarks) with explicit diquark degree of freedom using QCD sum rule~\cite{Kim2011} have been discussed.
The description of the heavy baryon as heavy quark/diquark ($Q\phi$) using the BS equation will be investigated as our future work.

The tetra-quark states $qq\bar{q}\bar{q}$ may include diquark/antidiquark components. Although the two mesons molecular states may dominate in the tetra-quark due to the strong correlation between quark and antiquark, the diquark/antidiquark would be also important components~\cite{Maiani2005,Ding2006,Zhang2007,Lee2006,Heupel2012}. 
The tetra-quark states would be described as the linear combination of two mesons and diquark/antidiquark states based on the BS formalism.
The structure of sigma meson (light scalar mesons) is also applicable subject. The sigma meson is considered as a chiral partner of the pion in the context of the chiral symmetry, which structure is quark/antiquark bound state. The possibility of the light scalar mesons as four-quark states have been discussed~\cite{Jaffe1977,Zhang2007,Lee2006,Heupel2012,Alford2000,Suganuma2007,Prelovsek2010,Wakayama2012,Alexandrou2013,Chen2007,Chen2010,Sugiyama2007,Fariborz2008,Hooft2008}. The structure of the sigma meson (light scalar mesons) can be described as the linear combination of quark/antiquark, diquark/antidiquark and two mesons in the context of the BS formalism.

\section*{\color{myaqua} Acknowledgment}
S.I. thanks T.M. Doi, H. Iida and N. Yamanaka 
for useful discussion and comments. 
This work is in part supported by the Grant for Scientific Research 
[Priority Areas ``New Hadrons'' (E01:21105006), (C) No.23540306,  No.15K05076] 
from the Ministry of Education, Culture, Science and Technology of Japan

{\color{myaqua}

}

\end{document}